\documentclass[trackchanges]{aastex7}  

\usepackage{lineno}
\usepackage{makecell}
\usepackage{braket}
\usepackage{euscript}
\usepackage{hyperref}

\begin{document}
\linenumbers

\title{Dynamics of Subsurface Flows in Solar Active Regions During the May 2024 Storm}

\author[0000-0003-0666-7650]{B. Lekshmi}
\email{lekshmib@nso.edu}
\affiliation{National Solar Observatory, 3665 Discovery Dr., Boulder, CO 80303, USA }

\author[0000-0002-4995-6180]{Sushanta Tripathy}
\email{stripathy@nso.edu}
\affiliation{National Solar Observatory, 3665 Discovery Dr., Boulder, CO 80303, USA }

\author[0000-0002-1905-1639]{Kiran Jain}
\email{kjain@nso.edu}
\affiliation{National Solar Observatory, 3665 Discovery Dr., Boulder, CO 80303, USA }

\author[0000-0003-0489-0920]{Alexei Pevtsov}
\email{apevtsov@nso.edu}
\affiliation{National Solar Observatory, 3665 Discovery Dr., Boulder, CO 80303, USA }




\begin{abstract}
In May 2024, the Sun exhibited intense magnetic activity, marked by numerous high-intensity flares resulting from the interaction and merging of NOAA ARs 13664 and 13668 in the southern hemisphere and AR 13663 in the northern hemisphere. Notably, AR 13664 displayed an extended lifetime, remaining visible after a full solar rotation and continuing to produce significant flaring activity. In this study, we investigate the evolution of sub-photospheric plasma flows associated with these ARs during their disk passage using ring-diagram analysis of SDO/HMI Dopplergrams. We analyze flow divergence, vorticity, and kinetic helicity across depths from the surface to 25 Mm, revealing pronounced temporal and depth-dependent variations. Our observations indicate that the majority of flares occur on the days when the Normalized Helicity Gradient Variance, a measure of kinetic helicity spread, peaks or on the following day. Furthermore, we examine the relationship between subsurface flow dynamics and surface magnetic properties of these complex active regions to understand the interaction between them.
\end{abstract}


\section{Introduction} \label{sec:intro}
Active regions (ARs), characterized by strong magnetic fields on the solar surface, are key drivers of high-energetic solar events such as flares and Coronal Mass Ejections (CMEs). They provide direct, observable evidence of the critical role magnetic fields play in solar interior dynamics. In the convection zone, the kinetic helicity of turbulent plasma flows twists the rising flux tubes, a process known as the $\Sigma$-effect \citep{Longcope1998}. Upon emerging as ARs, surface shearing motions and rotational flows further energize and twist the overlying magnetic loops \citep{Vemareddy2012}. The accumulation of magnetic energy and non-potentiality in these flux tubes makes them highly unstable \citep{Romano2005}. When these unstable structures reach the solar atmosphere, they can explosively release stored energy, triggering solar flares and CMEs. 

While this idealized model highlights the role of plasma flows in destabilizing flux tubes and triggering eruptions, a more comprehensive understanding of the underlying mechanisms is essential for improving the accuracy of solar eruption forecasts. The complex interplay between evolving magnetic fields and plasma flows is central to driving such energetic events. Therefore, identifying dynamic parameters that capture the temporal evolution of both magnetic and kinetic properties is of paramount importance \citep{Toriumi2019}. Numerous studies have explored the evolution and interrelation of the kinetic and magnetic parameters in ARs, aiming to better understand their role in flaring activity. However, earlier studies using the vertical component of current helicity as a proxy for magnetic helicity \citep{Gao2009} and the force-free field parameter derived from photospheric magnetic fields \citep{Maurya2011} failed to establish a clear relationship between the kinetic and magnetic helicities of ARs. Analyzing long-lived activity complexes, \cite{Komm2015} identified a consistent sign correlation between kinetic and magnetic helicities. However, a recent study by \cite{Liu2024} found no significant correlation between kinetic helicity and any magnetic parameters.  The lack of correlation may be attributed to the broader, less targeted approach of their analysis involving flow measurements derived from standard Helioseismic and Magnetic Imager \citep[HMI:][]{HMI2012} ring–diagram pipeline \citep{HMI_Ring} rather than being specifically targeted on the selected ARs. In contrast, a custom analysis of ARs producing high-intensity flares revealed correlations between magnetic and subsurface kinetic properties \citep{Lekshmi2022}.  The study also found that integrated flow and magnetic parameters measured one day prior to the onset of flares are in-tandem with the total flare intensity.  However, similar to many earlier studies, this result is based on ensemble averages and not on individual ARs.

\cite{Komm2009} have observed that the intensity of flares can be characterized by the values of vorticity and magnetic flux. The unsigned zonal and meridional vorticity components of ARs have been found to correlate with total flare intensity, providing additional predictive insights \citep{Mason2006}. Using the technique of helioseismic holography, \cite {Barun2016} also studied a few flow parameters and found modest correlation with integrated flare X-ray flux, qualitatively similar to \cite{Komm2009}. However, the analysis did not reveal any distinct signature of temporal variations or precursors associated with flares. Investigations of solar flare precursors suggest that the spread of kinetic helicity in ARs, quantified as the Normalized Helicity Gradient Variance (NHGV), peaks prior to high-intensity flares. The temporal evolution of  NHGV  exhibits a distinct peak, occurring 2–3 days before high-intensity flares in ensemble-averaged AR studies \citep{Reinard2010} or around the flare onset in a few individual ARs \citep{Gao2014}.

Solar cycle 25 has generated numerous X-class flares offering a new opportunity to explore the connections between magnetic and sub-surface flow properties of the ARs. In this paper, we analyze the subsurface flow dynamics of ARs associated with the intense solar storm of 2024 May, one of the most powerful geomagnetic events in recent history and examine potential pre-flare signatures in the flow parameters. We also explore the correlations between the flow parameters of each of these ARs in the outer 3\% of the Sun below the surface and the photospheric magnetic field parameters. 

\begin{figure}[h]
    \centering  \includegraphics[width=0.5\linewidth]{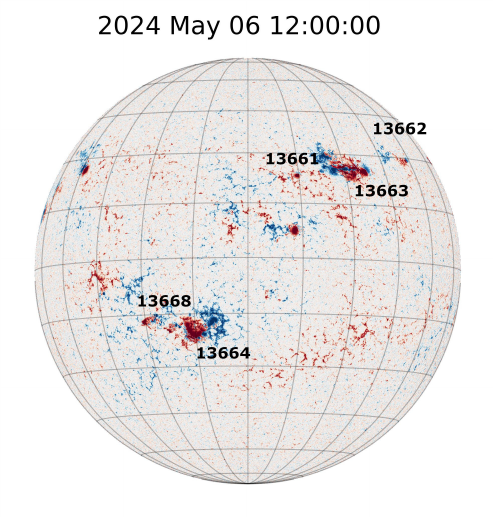}  
    \caption{HMI full disk magnetogram on 2024 May 06 12:00:00. The NOAA ARs 13661, 13662, 13663, 13664 and 13668 are indicated.}
    \label{fig:magnetogram}
\end{figure}

\section{Analysis} \label{sec: analysis}

\subsection{Data and Technique}

In 2024 May, the Sun exhibited heightened activity with the emergence of multiple ARs, several of which produced high-intensity flares. Among these, NOAA AR\,13664 was particularly significant, generating multiple major eruptions.  Numerous studies have already reported on the magnetic properties of these ARs and the interplanetary impacts of their associated eruptions \citep{Jarolim2024, Jaswal2025, Hayakawa2025, Wang2024, Khuntia2025, Ippolito2025}. In this investigation, in addition to AR\,13664, we extend our analysis to ARs\,13663 and 13697. AR\,13663 was located in the northern hemisphere and also produced multiple high intensity flares, while AR\,13697 was formerly identified as AR\,13664 in the previous rotation. We focus on the temporal evolution of subsurface kinetic properties of these ARs and their relationship with corresponding magnetic parameters. All three regions offer extended coverage, enabling a prolonged observation period as they traverse the solar disk.

Both ARs\,13663 and 13664 belong to the Hale class $\beta\gamma\delta$ and produced numerous X- and M-class flares, with the most intense events being an X4.5 flare from AR\,13663 on 2024 May 6 and an X8.8 flare from AR\,13664 on 2024 May 14. AR\,13697 also belongs to the class $\beta\gamma\delta$ and produced its most intense flares, with an X1.4 intensity on 2024 May 29 and another flare of same intensity on 2024 Jun 01. Details of X- and M-class flares produced by these regions during the observation period used in this work are tabulated in Appendix~\ref{App: table}.

We define region AR 13663 as a complex comprising NOAA ARs\,13661, 13662, and 13663, while region AR\,13664 includes both ARs\,13664 and 13668. These complexes correspond to HMI AR Patches (HARPs) 11142 and 11149, respectively. 
The HMI full-disk magnetogram on 2024 May 06 at 12:00:00 UT, highlighting these ARs is included in Figure~\ref{fig:magnetogram}. According to HMI HARP catalog, AR\,13663 reached a maximum area of 1813.52 $\mu Hemi$, while AR\,13664 extended to 4361.79 $\mu Hemi$. In comparison, AR\,13697 had a maximum area of 3836.41 $\mu Hemi$, smaller than that of AR 13664, suggesting that the region was already undergoing decay. Moreover, AR\,13664 is a complex region consisting of two large ARs while AR\,13697 contains only a single AR. 

\begin{figure}[h]
    \centering  \includegraphics[width=0.8\linewidth, trim=80 80 80 20, clip]{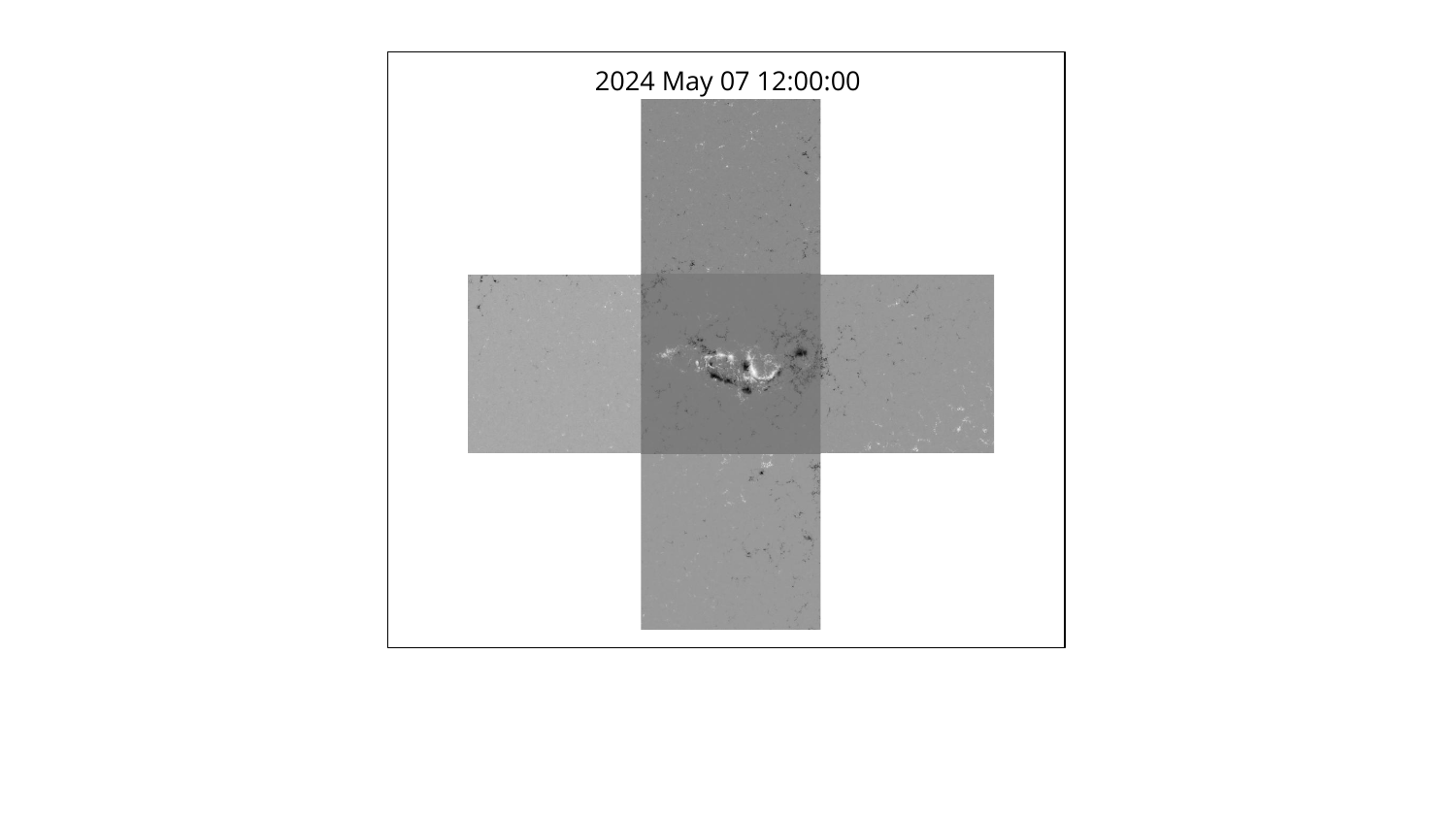}  
    \caption{AR 13664 and its neighboring patches on 2024 May 07 12:00:00 UT. The flow vectors are computed for all the patches shown here. To compute the flow gradients, we use the flow vectors from the adjacent patches surrounding the active region.}
    \label{fig:patch_selec}
\end{figure}

To determine horizontal flow velocities in the subsurface layers, we apply ring-diagram analysis \citep[RDA:][]{Hill1988, Corbard2003} to the Doppler velocity measurements from HMI onboard the Solar Dynamics Observatory \citep[SDO:][]{SDO2012} recorded at a 45-second cadence with one arc-second resolution. Since the ring-diagram technique uses three-dimensional spectra from  small patches to calculate flows below the surface, we selected square patches on the solar disk ensuring that the entire AR is contained within the patch \citep{Jain2015}. Additional patches of the same size, placed around each AR patch on all sides, are also processed through the RDA so that the gradient in flow parameters over latitude and longitude can be computed. The patch sizes vary for each AR and are listed in Table~\ref{tab:AR_info}, along with the location of the AR patch on each day. This selection of size ensures that the measured flow vector corresponds to the flow within the entire AR, and the adjacent patches do not contain fragments of the AR.
It also enables a direct comparison between the flow parameters and the SHARP magnetic parameters. Furthermore, the chosen patch sizes are large enough to allow the study of flow parameters down to a depth of 25 Mm below the surface, which is not feasible with the standard $15^{\circ}$ patches. These selected regions are tracked for 24-hrs using the surface rotation rate \citep{Snodgrass1984} as they move across the solar disk within a central meridian distance (CMD) of $\pm 45^{\circ}$. Thus, the flows obtained from RDA represent daily averages. AR 13664 and its adjacent patches as observed on 2024 May 07 12:00:00 UT are shown in Figure \ref{fig:patch_selec}.
\begin{deluxetable*}{lcccrccr}
\tabletypesize{\scriptsize}
\tablewidth{0pt}
\tablecaption{Size and location of AR patches along with quiet periods. \label{tab:AR_info}}
\tablehead{
\colhead{AR} & \colhead{Patch size} & \colhead{Day \#} & {Date} & \multicolumn{2}{c}{Location\textsuperscript{a,b}} & {Quiet Period\textsuperscript{c}}\\
\cline{5-6}
    \colhead{} & \colhead{} & \colhead{} & \colhead{} & \colhead{Long $(\phi_{\text{AR}})$} & \colhead{Lat $(\theta_{\text{AR}})$}& 
}
\startdata
13663 & $25.92^{\circ}$&&&&$25.6^{\circ}$&
\\
& & \makecell{1 \\ 2\\ 3\\ 4\\ 5\\ 6\\ 7} & \makecell{2024 Apr 30  \\ 2024 May 01\\ 2024 May 02 \\ 2024 May 03 \\ 2024 May 04\\ 2024 May 05 \\ 2024 May 06} & \makecell{$-43.7^{\circ}$ \\ $-30.5^{\circ}$ \\ $-17.3^{\circ}$ \\ $-4.1^{\circ}$ \\ $9.1^{\circ}$ \\ $22.3^{\circ}$ \\ $35.5^{\circ}$} & &  \\
\\
 & & \makecell{1 \\ 2\\ 3\\ 4}& \makecell{-\\-\\-\\-}& \makecell{-\\-\\-\\-} & \makecell{-\\-\\-\\-} &\makecell{2024 Apr 27  \\ 2024 Apr 28 \\ 2024 May 23 \\ 2024 May 24} & \\
 \hline 
 13664 & $27.2^{\circ}$&&&&$-18.9^{\circ}$&
 \\
 & &\makecell{1 \\ 2\\ 3\\ 4\\ 5\\ 6\\ 7} & \makecell{2024 May 04  \\ 2024 May 05 \\ 2024 May 06 \\ 2024 May 07 \\ 2024 May 08  \\ 2024 May 09  \\ 2024 May 10} & \makecell{$-40.6^{\circ}$ \\ $-27.4^{\circ}$ \\ $-14.2^{\circ}$ \\ $-0.98^{\circ}$} & &\\ 
 \\
 & & \makecell{1 \\ 2\\ 3\\ 4}& \makecell{-\\-\\-\\-}& \makecell{-\\-\\-\\-} & \makecell{-\\-\\-\\-} 
 & \makecell{2024 Apr 01  \\ 2024 Apr 02 \\ 2024 May 27  \\ 2024 May 28 } & \\
 \hline 
 13697 & $27.2^{\circ}$ &&&& $-18.9^{\circ}$&
 \\
 &&\makecell{1 \\ 2\\ 3\\ 4\\ 5\\ 6\\ 7} & \makecell{2024 May 31  \\ 2024 Jun 01 \\ 2024 Jun 02  \\ 2024 Jun 03  \\ 2024 Jun 04  \\ 2024 Jun 05 \\ 2024 Jun 06} & \makecell{$-40.6^{\circ}$ \\ $-27.4^{\circ}$ \\ $-14.2^{\circ}$ \\ $-0.98^{\circ}$ \\ $12.2^{\circ}$ \\ $25.4^{\circ}$ \\ $38.6^{\circ}$} & &\\
 \\
 & &  \makecell{1 \\ 2\\ 3\\ 4}& \makecell{-\\-\\-\\-}& \makecell{-\\-\\-\\-} & \makecell{-\\-\\-\\-} 
 & \makecell{2024 Apr 01  \\ 2024 Apr 02 \\ 2024 May 27  \\ 2024 May 28 }\\
 \enddata
\tablecomments{\\a: Locations are taken from HMI SHARP database \\
b: The neighboring patches of the ARs are positioned as $(\phi_{\text{AR}} \pm \text{Patch size}, \theta_{\text{AR}})$ and $(\phi_{\text{AR}}, \theta_{\text{AR}} \pm \text{Patch size})$. 
\\c: For each quiet period, patches are selected across all locations spanned by the AR and its surrounding neighbors throughout the observing period.}
\end{deluxetable*}

\subsection{Systematics Correction}
The measured flow velocities are known to be affected by systematic errors, such as the center-to-limb effect \citep{Zhao2012} and variations in the $B_0$ angle \citep{Zaatri2006}. The center-to-limb effect increases with distance from the disk center, while the $B_0$ angle variation results from the annual change in the Sun's inclination relative to Earth. To accurately analyze flow variations, it is essential to correct for these systematics. Various methods have been proposed in the literature to address these effects \citep[e.g.,][]{Komm2015a}. However, most of these corrections are applicable while studying long-term flow variations over multiple Carrington rotation time scales. 

\begin{figure}
    \centering
    \includegraphics[width=0.7\linewidth]{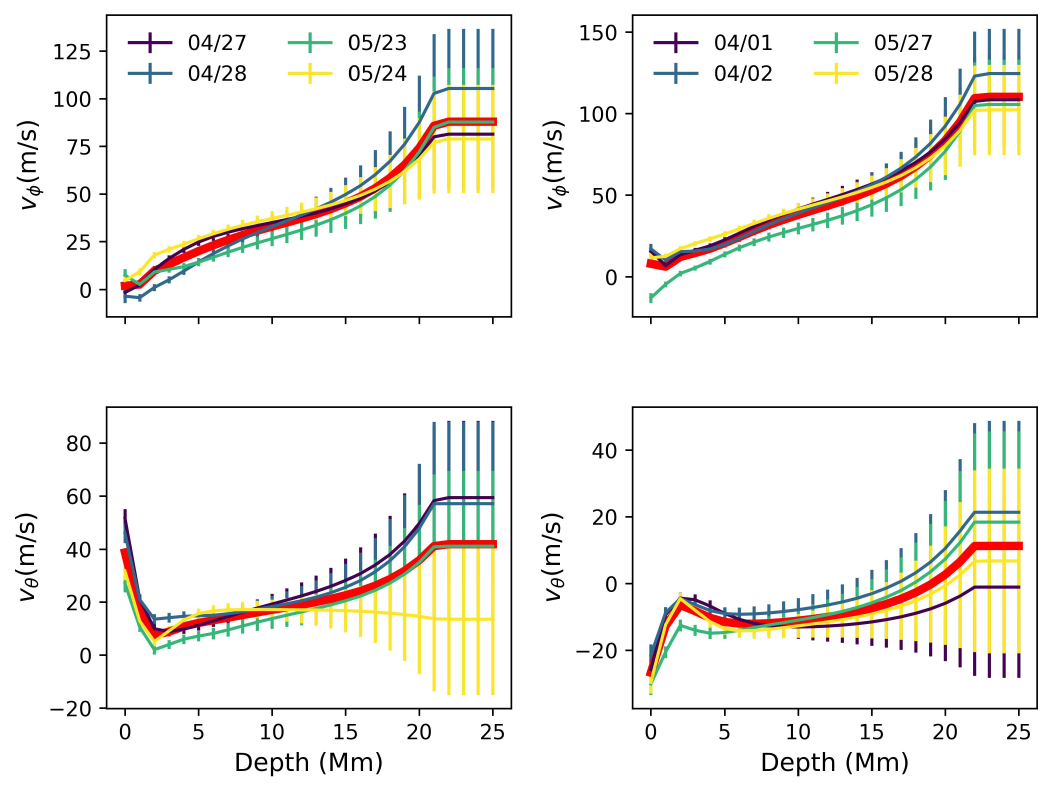}  
    \caption{The $\rm{v_\phi}$ (top) and $\rm{v_\theta}$ (bottom) components of quiet regions corresponding to ARs 13663 (left) and 13664/13697 (right) when the patch centers were at central meridian distances of $-4.11^{\circ}$ and $-0.98^{\circ}$, respectively. 
    The red solid line represents the error weighted mean of these velocities.}
    \label{fig:control_region_velocities}
\end{figure}

Since our focus is on flow variations over a few days, we adopt an approach similar to \citet{Jain2015}, where the velocities of quiet regions (QRs) from nearby Carrington rotations are subtracted from the measured flows to remove systematic effects. To do this, we identify days from the same, previous or following Carrington rotations with no solar activity at any of the locations covered by the AR and its neighboring patches throughout the entire tracking period. As a result, each AR patch is associated with four QRs. The selected quiet Sun periods for each AR are listed in Table~\ref{tab:AR_info}. Since AR 13664 is renumbered as AR 13697 in the  subsequent Carrington rotation, we use the same QRs for both. For example, Figure~\ref{fig:control_region_velocities} displays the velocities of the QR patches as they cross the central meridian. To correct for systematics, we subtract the error-weighted mean of the QR velocities (red solid line in Figure~\ref{fig:control_region_velocities}) from the corresponding AR and neighboring patches.

\subsection{Subsurface Flow and Magnetic Parameters}
Using the corrected horizontal flow velocities, we compute the divergence ($\nabla \cdot v_h$) and vorticity ($\nabla \times v_h$) of the ARs in spherical coordinates as, 
\begin{equation}
    \nabla.v_h = \frac{1}{r\sin\theta}\left (\frac{\partial \rm{v_\phi}}{\partial \phi} + \frac{\partial}{\partial \theta}(\sin\theta \rm{v_{\theta}})\right),
    \label{eq.divergence}
\end{equation}
\begin{equation}
    \nabla \times v_h = \frac{1}{r\sin\theta}\left (\frac{\partial}{\partial\theta} (\sin\theta \rm{v_\phi}) + \frac{\partial \rm{v_\theta}}{\partial\phi}\right),
    \label{eq:vorticity}
\end{equation}
where $\phi$ and $\theta$ represent the longitude and co-latitude, $\rm{v_\phi}$ and $\rm{v_\theta}$ are daily averages of the zonal and meridional velocities and $v_h$ is the horizontal velocity vector ($\rm{v_\phi}, \rm{v_\theta}$). The spatial derivatives in Equations \ref{eq.divergence} and \ref{eq:vorticity} are computed using the finite difference method. Specifically, we apply the central difference scheme using the flow velocities from the four non-overlapping patches surrounding the active region patch.

Using the continuity equation, the radial component of velocity can be computed as \citep{Komm2019},
\begin{equation}
    v_r(r') = -\frac{1}{\rho(r')}\int_{R_{\odot}}^{r'}\rho (\nabla.v_h)dr + v_r(R_\odot)\frac{\rho(R_\odot)}{\rho(r')}
    \label{eq:vr} ,
\end{equation}
at depth $d$ with $r' = R_\odot - d$, where $v_r(R_\odot)$ and $\rho(R_\odot)$ are the radial velocity and density at the solar surface. We use the density $\rho$ from the solar model S of  \cite{Dalsgaard1998}. The radial component of helicity scalar $k_r$ is defined as the product of the radial components of vorticity and velocity,
\begin{equation}
    \braket{k_r} = \braket{(\nabla \times v_h) \cdot v_r} 
\end{equation}
Finally, the kinetic helicity in the radial direction is computed by integrating $k_r$ in volume as,
\begin{equation}
    K_r = \int_{R_\odot - d_1}^{R_\odot -d_2}k_r(r)A_p(r)dr,
    \label{eq:kinetichelicity} ,
\end{equation}
where $d_1$ and $d_2$ are the depth range and $A_p$ is the area of the patch chosen to compute the velocities. 
The temporal helicity change at each depth (r) from one day to the next is,
\begin{equation}
     \Delta k_r(t) = k_r(t) - k_r(t-1).
\end{equation}
Then the spread in temporal helicity change with depth can be calculated as,
   \begin{equation}
       \Delta k(t) = \sum_r (\Delta k_r(t) - \Delta k_{r+1}(t)),
   \end{equation}
   where $r$ takes only the odd indices of depth (1, 3, 5, \ldots). This quantity reflects the extent to which the evolution of kinetic helicity varies across depth at a given time. The vertical gradient of kinetic helicity is calculated as, 
\begin{equation}
    k(t) = \sum_r (k_r(t) - k_{r+1}(t))
\end{equation}
The product of $\Delta k(t)$ and $k(t)$ gives the  Normalized Helicity Gradient Variance (NHGV; \citealt{Reinard2010}),
\begin{equation}
    \mathrm{NHGV} = \Delta k(t)\; k(t),
    \label{eq:NHGV}
\end{equation}
A large value of $k(t)$ indicates a strong vertical gradient in kinetic helicity, while a large value of $\Delta k(t)$ suggests that the helicity at different depths is evolving at different rates. Thus, NHGV serves as a composite parameter that captures both the spatial structure and temporal evolution of kinetic helicity beneath active regions and characterizes the complexity of the sub-surface flows associated with flare producing regions.

The magnetic parameters, unsigned magnetic flux, current helicity, and magnetic twist used in this study are obtained from HMI-SHARP data base  \citep{Bobra2014}. We compute a proxy for magnetic writhe ($W$) in the ARs, as $W = \lambda/d$, where $\lambda$ is the tilt angle of the AR in radians and $d$ is the separation in centimeters \citep{Liu2024} derived from the centroid locations of the positive and negative polarities. The parameters are averaged over the same time interval to match the temporal resolution of measured flows.
\begin{figure}
    \centering
    \includegraphics[width=0.98\linewidth]{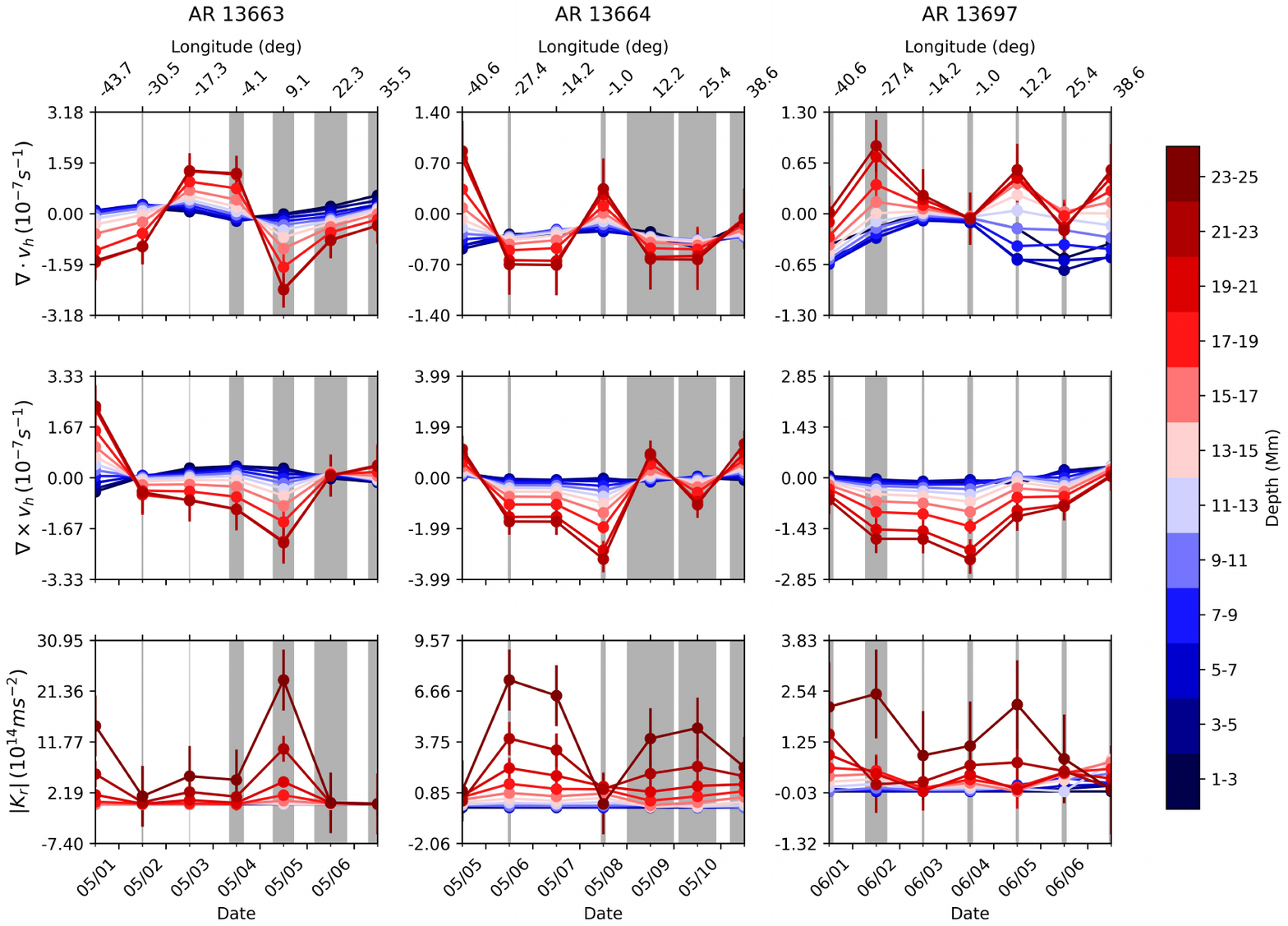}  
    \caption{The divergence ($\nabla \cdot v_h$), vorticity ($\nabla \times v_h$), and absolute value of kinetic helicity ($K_r$) averaged over different depth ranges of the ARs along with errors are plotted as a function of time. 
    The grey vertical bars represent the weighted sum of flares on each day, where a weight of 0.5 is assigned to an M1-class flare, increasing linearly by 0.5 for each subsequent class, with an X4-class flare assigned a weight of 6.5. To maintain consistency with the velocity measurements, the bars are centered at 12:00:00 UT each day.}
    
    \label{fig:flow_parameters}
\end{figure}

\section{Results} \label{sec:results}
\subsection{Temporal Variations in Flow Parameters}
The divergence, vorticity and absolute value of kinetic helicity ($|K_r|$) of the ARs over different depth slices are shown in Figure~\ref{fig:flow_parameters}. The divergence and vorticity of the flows increase with depth and exhibit large amplitudes primarily around the occurrence of high-intensity flares. Despite the reduction in size of AR\,13697, its divergence and vorticity remain comparable to its predecessor, AR\,13664. The persistence of strong internal dynamics even after one solar rotation suggests sustained sub-surface forcing due to the large size or the strength of the AR, delaying the expected decay of AR\,13664. The flow divergence is predominantly negative over all depths for ARs 13663 and 13664, indicating that the flows are mostly converging toward the AR. However, for AR\,13697, the flows exhibit mostly positive divergence at deeper depths, indicating predominant outflows, while the near-surface layers show converging flows. The vorticity of all three ARs are mostly negative, irrespective of the hemisphere in which they are located.

The absolute value of $K_r$ is significantly higher around the time periods associated with high intensity flares (Figure~\ref{fig:flow_parameters} bottom panel) and its amplitude increases with depth. AR\,13663 has a larger $|K_r|$ compared to the other ARs. It showed a steady increase from 01 May, reaching a strong peak on 04 May and aligning with numerous high-intensity flares that occurred during this period. AR\,13664 exhibits two peaks—one on 05 May, around which a few flares occurred, followed by a decline. The value then began rising again from 07 May, peaking on 09 May, coinciding with numerous high-intensity flares including many X-class flares. AR\,13697 exhibited peaks around 01 Jun, marked by the occurrence of multiple X-class flares, and again on 04 Jun, with additional flaring activity. During this period, $|K_r|$ was particularly strong at depths greater than 20\,Mm. It may also be noted that the amplitudes of kinetic helicity of QRs are 2~–~3 orders of magnitude smaller than those observed in ARs.

\subsection{ Temporal Variation of NHGV and Flare Precursor}

\begin{figure}
    \centering  \includegraphics[width=0.98\linewidth]{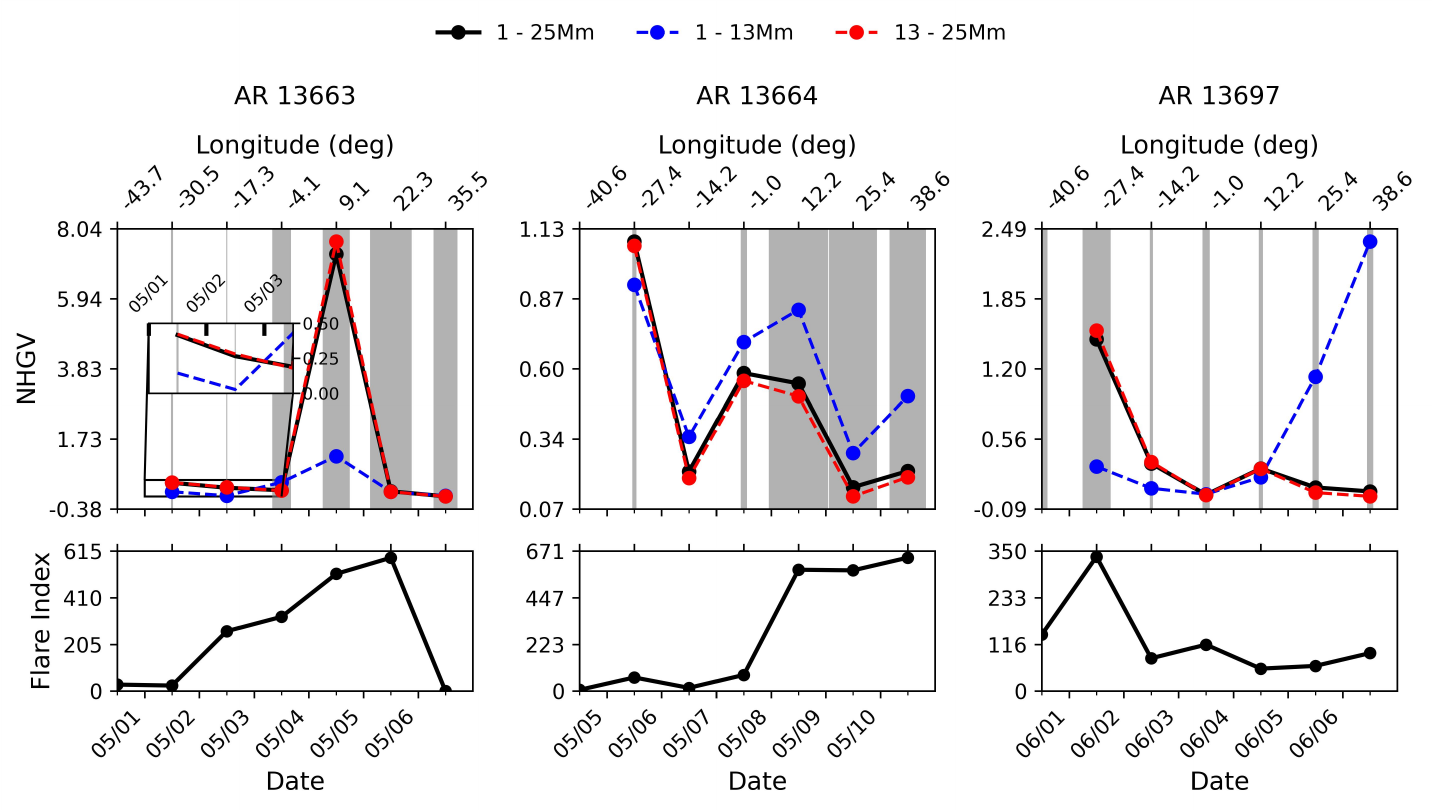}  
    \caption{ Top row: NHGV  computed using helicities over the depth ranges 1~-~25~Mm (black solid line), 1~-~13~Mm (blue dashed line), and 13~-~25~Mm (red dashed line). The values are normalized by the mean NHGV of QRs. The inset in the  NHGV  plot of AR 13663 provides a zoomed-in view of values between 01 May 2024 and 03 May 2024.
    The vertical bars in the background represents the weighted sum of number of flares on each day, same as in Figure~\ref{fig:flow_parameters}. Bottom row: Daily flare index of each AR.}
    \label{fig:NHGV}
\end{figure}

We compute  NHGV  of the ARs using kinetic helicity measurements over the depth range of 1~–~25\,Mm as well as within two specific depth intervals: 1~-~13\,Mm and 13~–~25\,Mm. The NHGV computed for each AR using Eq.~\ref{eq:NHGV} is normalized by the mean of QRs and are shown in top row of Figure~\ref{fig:NHGV}. Bottom row presents the flare index of ARs during the tracking period computed as $\Sigma 100 \times I_X + 10 \times I_M + I_C$, where $I_X$, $I_M$ and $I_C$ represent the GOES magnitude of X-, M- and C-class flares respectively \citep{Sun2015}. This provides a measure of the daily flaring activity level of the ARs.

The NHGV of AR 13663 measured over 1~–~25\,Mm and 13~–~25\,Mm are similar, whereas those over 1~–~13\,Mm are smaller. The inset figure shows that the NHGV on the first day of observation is higher compared to the following two days. This peak and the subsequent decline in NHGV, are accompanied by multiple flares. Following this, the NHGV sharply increases, peaking on 04 May, with numerous flares occurring at the peak and during its decline on 05 May. 

AR 13664 exhibits multiple peaks in NHGV throughout its disk passage, with a pronounced peak occurring on 05 May across all depth ranges followed by peaks on 07 and 08 May and a rise  on 10 May. Similar to AR 13663, AR 13664 also generated numerous flares around the peaks and during the decline of NHGV. The variation in NHGV remains consistent across all depth ranges, with the 1–13\,Mm range exhibiting higher values starting from 06 May and peaking on 08 May. This region, while near the limb produced multiple M and X-class flares between 10 and 14 May including an X8.8-class flare on 14 May. Although this period falls outside of our analysis window, due to the limitation imposed by ring-diagram technique, the observed rising trend in NHGV from a local minimum on 09 May probably indicates that a substantial increase in NHGV likely occurred prior to these events. 

AR 13697 displays two distinct peaks in NHGV values with the first peak on 01 June and the second on 04 June , both accompanied by flaring activity. The values over the 13~-~25\,Mm range are similar to those over 1~-~25\,Mm. Although the variation in NHGV computed over 1~-~13\,Mm generally follows the trend observed in the other two cases, with an exception of a sudden rise after 04 May.

A key finding from this analysis is that the NHGV begins to rise at least a day prior to all major flaring events, characterized by high flare index values. The temporal evolution of NHGV is largely in phase with the evolution of the flare index of the ARs. All the observed peaks in NHGV coincides with major flares (grey vertical bars in Figure~\ref{fig:NHGV}). Also, numerous flares are observed during its subsequent decline. Quantitatively, 49\% of the flares observed across all ARs occurred on the day when the NHGV reached a peak, and 32\% were observed on the following day. Thus majority of flares (81\%) occurred either on the day of or the day after the NHGV attained a local maximum. It is promising to find that we were able to capture NHGV peaks even around comparatively low-intensity flares on 07 May and 04 June. These observations supports the potential use of NHGV as a flare precursor, as previously suggested by \cite{Reinard2010} and \cite{Gao2014}.

\subsection{Relationship between Kinetic and Magnetic Parameters}

In order to examine the relationship between kinetic and magnetic parameters, we compute these  parameters over shallower (1~-~13\,Mm) and deeper (13~-~25\,Mm) layers, consistent with the depth ranges chosen for NHGV calculations. We combine the parameters of all three ARs together to improve the statistical significance of the correlation analysis. 

\begin{figure}
    \centering  \includegraphics[width=0.98\linewidth]{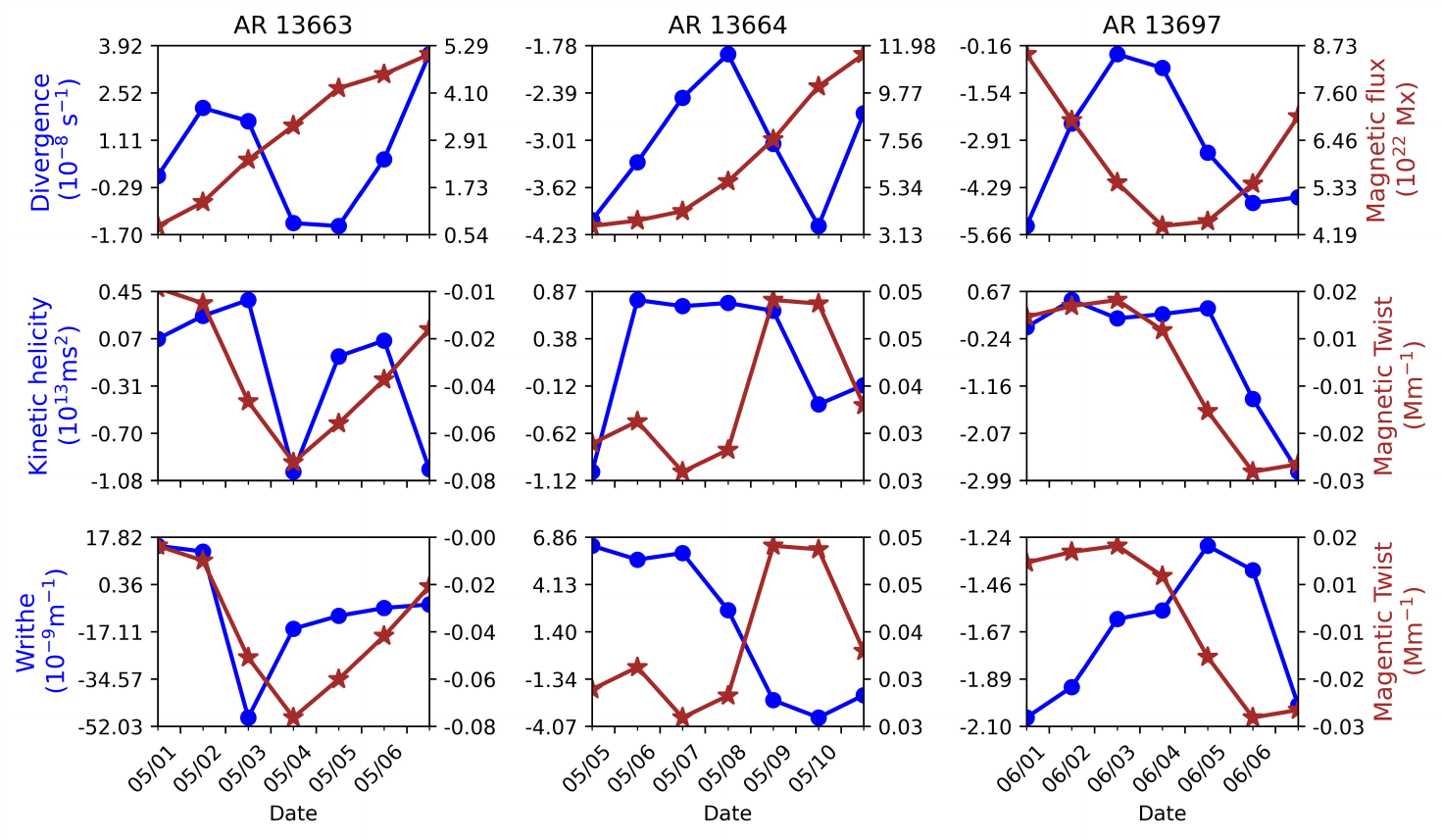}  
    \caption{The kinetic and magnetic parameters of ARs as function of time. Top: The temporal evolution of AR horizontal flow divergence averaged over depths 1~-~13~Mm (blue) and unsigned unsigned magnetic flux (brown),  
    Middle: The kinetic helicity of ARs averaged over depths 1~-~13~Mm (blue) and magnetic twist parameter (brown), 
    Bottom: The AR magnetic writhe (blue) and magnetic twist (brown). }
    \label{fig:mag_flow_param}
\end{figure}

The top panel of Figure~\ref{fig:mag_flow_param} shows the divergence averaged over shallower depths and unsigned magnetic flux as a function of time. We observe that the divergence is mostly negative for large values of magnetic flux. A moderate Pearson correlation coefficient $r_p$ =  $-0.52$ (99\,\% confidence) is obtained between the two parameters implying a convergent flow towards regions of stronger magnetic activity. We do not observe any significant correlation for the deeper layers, hence the divergence and other parameters are only shown over the depth range 1~-~13\,Mm in Figure~\ref{fig:mag_flow_param}. 

The middle panel of Figure~\ref{fig:mag_flow_param} shows the temporal evolution of kinetic helicities and the magnetic twists, proxy for current helicity \citep{Pevtsov1995}. The parameters are mostly in phase for ARs 13663 and 13697, while it shows deviation for AR 13664. We observe that the combined kinetic helicities of all ARs yield a Pearson correlation coefficient of 0.42 (97\,\%) with the magnetic twist, indicating a weak linear correlation between the two parameters. \cite{Lekshmi2022} also reported a linear correlation between the kinetic and current helicities of flaring ARs, consistent with our findings. 
The kinetic helicities averaged over deeper layers do not show any significant correlation with twist. 
Although the observed near-surface correlation is weak, likely due to poor statistics, it suggests that the current helicity is either generated by the twisting and writhing of flux tubes induced by subsurface flows, or represents the response of these flows to the transport of magnetic helicity from the lower to the upper parts of the flux tubes. To better understand this, we examine the relations between the magnetic writhe and twist of the ARs.

The magnetic writhe and twist of the ARs are plotted as a function of time in the bottom panel of Figure~\ref{fig:mag_flow_param}. These quantities are mostly in phase for ARs 13663 and 13697, but show a noticeable deviation for AR 13664, similar to the behavior observed in the kinetic helicity and twist. A linear correlation is observed between the writhe and twist, with a moderate Pearson correlation coefficient of 0.52 (99\,\%) in agreement with \cite{tian2003} who studied a sample of 86 major flare-producing ARs. According to the principle of helicity conservation, twist and writhe should have opposite signs if the initial twist of the flux tube prior to emergence is very small or negligible. Therefore, the observed same-sign relationship from our analysis implies that the flux tube emerged with significant initial twist. This, in turn, suggests that the observed correlation between twist and kinetic helicity reflects the response of subsurface flows to the pre-existing magnetic helicity of the active region.

 
\section{Discussion}
We examined the time-dependent variations of subsurface flow parameters beneath the ARs associated with the 2024 May solar storm. In particular, we focused on ARs\,13663, 13664 and 13697 which are large and complex ARs. The observed flow divergence and vertical vorticity are of the same order as reported in earlier studies \citep{Komm2004}. The amplitudes of these parameters are also found to be significantly stronger, at least an order of magnitude larger than those observed in quiet-Sun regions, consistent with \cite{Barun2016}.

We observe significant variations in all three flow parameters - divergence, vorticity, and NHGV across both temporal and depth scales. The amplitudes of these parameters tend to decrease following a series of high-intensity flares and again increases gradually before subsequent eruptions. \citet{Komm2009} reported that flare-prolific ARs exhibit large values of magnetic flux and vorticity. This is consistent with our observations, where pronounced peaks in vorticity are seen during time periods corresponding to sequences of high-intensity flares. 

\citet{Reinard2010}, using flow measurements from the GONG ring-diagram pipeline, reported that the NHGV increases 2–3 days prior to flares. Similarly, \citet{Gao2014}, using the time-distance helioseismology method, observed distinct bumps in NHGV within 8 hours of flare occurrences, with major flares associated with pronounced variations in this parameter. Using our custom measurements of AR flows, we observe that NHGV values tend to rise prior to major flares, and that the majority of flares produced by the ARs in this study occurred either around the local maximum of NHGV or on the following day. These observations reinforce the potential of NHGV as a predictive parameter for flare forecasting by identifying periods during which NHGV begins to rise.


Our analysis shows that the near-surface flow divergence (1~-~13\,Mm) is anti-correlated with the surface magnetic flux of the ARs, which agrees with the earlier observations of plasma flows converging toward regions of strong magnetic activity \citep{Gizon2004}. The kinetic helicity of ARs, averaged over depths of 1~–~13\,Mm, exhibits a positive correlation with the mean magnetic twist at the photosphere, a proxy for current helicity. Previous studies have also reported correlations between surface current helicity and subsurface kinetic helicity \citep{Gao2012, Lekshmi2022}. In the context of the $\Sigma$-effect, the writhing and twisting of magnetic field lines by subsurface flows contribute to the buildup of magnetic and current helicities in ARs  \citep{Longcope1998}. However, the observed linear correlation between twist and writhe suggests that flux tubes are already twisted before emerging at the surface, likely due to dynamo action. \citet{tian2003} studied the relationship between twist and writhe in major flaring active regions with $\beta\gamma\delta$ configuration and found that the majority (60\%) of regions exhibited the same sign for both parameters. They reported that if a flux tube emerges with significant twist and is highly kink-unstable, then both twist and writhe should have the same sign, as part of the twist is transferred to writhe during the development of the instability. This mechanism explains the same-sign relationship observed in our study for ARs with similar configuration. The numerous high-intensity flares produced by these regions imply they contain large amount of free energy resulting from highly twisted flux tubes. Consequently, the observed pre-flare increase in kinetic helicity/NHGV in these ARs may not result from flows generating twist prior to flare. Instead, it may reflect subsurface flows responding to the upward propagation of twist from the solar interior rather than subsurface flows injecting helicity into the magnetic field prior to eruption. 

In conclusion, we have investigated the subsurface flow properties of three major ARs from the current solar cycle. Our analysis reveals significant variations in flow parameters associated with these ARs around the times of high-intensity flares. Notably, we find that the spread in kinetic helicity, referred to as NHGV, begins to increase prior to flare onset, with most flares occurring near the peak in NHGV or shortly afterwards. Moreover, the weak but statistically significant correlations observed between kinetic parameters near the surface and magnetic parameters are promising results. To determine the precise timing of the NHGV rise relative to flare initiation and to quantitatively assess the potential of NHGV as a flare indicator, we plan a more comprehensive statistical analysis involving a larger sample of both flaring and non-flaring (quiet) complex active regions with higher temporal resolution. This will provide deeper insights into their physical connections and further probe internal dynamics of the ARs.

\begin{acknowledgments}
We thank the referee for the valuable comments. This work utilizes data from HMI onboard NASA's SDO spacecraft, courtesy of NASA/SDO and the HMI Science Teams. SCT and KJ acknowledge the partial support from NASA-DRIVE Center award 80NSSC20K0602  to Stanford, and NASA grant 80NSSC23K0404 to the National Solar Observatory. 
The National Solar Observatory (NSO) is operated by the Association of Universities for Research in Astronomy (AURA), Inc., under cooperative agreement with the National Science Foundation. 
\end{acknowledgments}


\facilities{SDO/HMI,GOES }

\bibliography{references}{}
\bibliographystyle{aasjournalv7}

\appendix

\section{List of flares} \label{App: table}

\begin{deluxetable*}{cc cc cc ccc}[h]
\tabletypesize{\scriptsize}
\tablewidth{0pt}
\tablecaption{X- and M-class flares observed during the study periods. \label{tab:flare_info}}
\tablehead{
\multicolumn{2}{c}{AR 13663} && \multicolumn{2}{c}{AR 13664} & &\multicolumn{2}{c}{AR 13697} \\
\cline{1-2} \cline {4-5} \cline{7-8}
\colhead{Date} & \colhead{Flare Class} & &
\colhead{Date} & \colhead{Flare Class} & &
\colhead{Date} & \colhead{Flare Class}
}
\startdata
\\
\makecell{2024 May 01 22:31:00\\ \\ 2024 May 02 02:17:00\\ \\ 2024 May 03 02:22:00\\  2024 May 03 08:11:00\\ 2024 May 03 23:16:00\\ 2024 May 03 23:30:00\\ \\ 2024 May 04 00:36:00\\ 2024 May 04 06:19:00\\ 2024 May 04 07:07:00\\ 2024 May 04 18:20:00\\ 2024 May 04 22:37:00\\ 2024 May 04 23:09:00\\  \\ 2024 May 05 01:28:00\\ 2024 May 05 05:59:00\\ 2024 May 05 08:19:00\\ 2024 May 05 09:56:00\\ 2024 May 05 11:54:00\\ 2024 May 05 14:47:00\\ 2024 May 05 15:38:00\\ 2024 May 05 19:52:00\\ \\ 2024 May 06 01:06:00\\ 2024 May 06 05:28:00\\ 2024 May 06 06:35:00\\ 2024 May 06 09:59:00\\ 2024 May 06 21:48:00\\ 2024 May 06 22:27:00\\ \\ \\ \\ \\ \\ \\  \\ \\ \\ \\ \\} & \makecell{M1.8\\ \\ M1.0\\ \\ X1.6\\ M4.4\\ M1.0\\ M2.4\\ \\ M1.6\\ M9.1\\ M1.5\\ M1.3\\ M3.2\\ M9.0\\ \\ M8.4\\ X1.3\\ M1.3\\ M7.4\\ X1.2\\ M1.3\\ M2.2\\  M1.3\\ \\ M1.6\\ M1.3\\ X4.5\\  M1.5\\ M1.2\\ M4.3 \\ \\ \\ \\  \\  \\ \\  \\ \\ \\ \\  \\} & & \makecell{2024 May 05 09:38:00 \\2024 May 05 17:00:00 \\2024 May 05 18:39:00 \\ \\2024 May 07 08:23:00 \\2024 May 07 11:47:00 \\2024 May 07 13:35:00 \\2024 May 07 20:22:00 \\ \\2024 May 08 02:27:00 \\2024 May 08 03:27:00 \\2024 May 08 05:09:00 \\2024 May 08 17:53:00 \\2024 May 08 19:21:00 \\2024 May 08 20:34:00 \\2024 May 08 21:40:00 \\2024 May 08 21:40:00 \\2024 May 08 22:27:00 \\ \\2024 May 09 03:17:00 \\2024 May 09 03:32:00 \\2024 May 09 04:49:00 \\2024 May 09 06:13:00 \\2024 May 09 09:05:00 \\2024 May 09 11:56:00 \\2024 May 09 12:12:00 \\2024 May 09 13:23:00 \\2024 May 09 17:44:00 \\2024 May 09 22:56:00  \\2024 May 09 23:51:00 \\ \\2024 May 10 00:13:00 \\2024 May 10 03:29:00 \\2024 May 10 06:54:00 \\2024 May 10 10:14:00 \\2024 May 10 14:11:00 \\2024 May 10 18:32:00 \\2024 May 10 18:48:00 \\2024 May 10 19:05:00 \\2024 May 10 19:53:00 \\2024 May 10 20:03:00 \\2024 May 10 21:08:00} & \makecell{M2.3 \\ M1.3\\ M1.0\\ \\ M1.3\\ M2.4\\ M1.0\\ M2.1\\ \\ M3.4\\ M1.8\\ X1.0\\ M7.9\\ M2.0\\ M1.7\\ X1.0\\ X1.0\\ M9.8\\ \\ M4.0\\ M4.5\\ M1.7\\ M2.3\\ X2.2\\ M3.1\\ M2.9\\ M3.7\\ X1.1\\ M1.2\\ M1.5\\ \\ M1.3\\ M1.4\\ X3.9\\ M2.2\\ M5.9\\ M1.1\\ M1.7\\ M2.0\\ M1.1\\ M1.9\\ M3.8} &
 & \makecell{2024 May 31 11:20:00  \\2024 May 31 22:01:00 \\  \\2024 Jun 01 08:48:00  \\2024 Jun 01 18:36:00  \\2024 Jun 01 19:39:00 \\  \\2024 Jun 02 04:50:00  \\2024 Jun 02 08:50:00  \\ \\2024 Jun 03 05:17:00  \\2024 Jun 03 11:55:00  \\2024 Jun 03 12:27:00  \\ \\2024 Jun 04 06:31:00  \\2024 Jun 04 09:04:00 \\  \\2024 Jun 05 08:56:00  \\2024 Jun 05 10:07:00  \\ \\2024 Jun 06 15:06:00 \\ \\ \\ \\  \\  \\ \\  \\ \\ \\ \\ \\ \\ \\ \\ \\ \\ \\ \\ \\ \\ \\} & \makecell{M1.0\\ X1.1\\ \\ X1.4\\ X1.0\\ M7.3\\ \\ M1.2\\ M2.0\\ \\ M1.0\\ M3.2\\ M2.8\\ \\ M2.4\\ M1.6\\ \\ M3.4\\ M2.6\\ \\ M6.1 \\ \\ \\ \\  \\  \\ \\  \\ \\ \\ \\ \\ \\ \\ \\ \\ \\  \\ \\ \\ \\ \\ } \\
\\
\enddata
\tablecomments{Data source: NOAA Space Weather Prediction Center reports \\    \url{https://www.swpc.noaa.gov/products/report-and-forecast-solar-and-geophysical-activity}}
\end{deluxetable*}
\newpage

\end{document}